# Bricks or Cash? Externalities of Housing Upgrading in High-density Cities[1]

Sumit Agarwal[a2] Ying Deng[b3] Yi Fan[a4] Qi Gao[c5] Jing Li[c6] Lin Ma[c7]

[a]*Department of Real Estate, National University of Singapore*

[b]*School of Economics, University of International Business and Economics*

[c]*School of Economics, Singapore Management University*

June 15, 2026

**Abstract**

We estimate housing externalities in a high-density city, exploiting the staggered rollout of Singapore's nationwide Main Upgrading Programme for public housing. Controlling for nonrandom neighborhood exposure, we find that upgrading raises treated buildings' prices by 11.5% upon completion and neighboring buildings' resale prices by about 2% within 500 meters, decaying to zero beyond. A model with distance-decaying externalities shows that in dense settings spillovers justify the distortions of in-kind provision; this advantage diminishes and reverses at lower densities. Administrative data on over 2 million residents show that upgrading disproportionately retains older incumbents, suggesting age-specific amenities as an underexplored externality channel.

[1]Our study does not generate new data. All data used are from secondary sources. This research/project was supported by the Singapore Ministry of Education (MOE) Academic Research Fund (AcRF) Tier 1 grant (Proposal ID: 24-SOE-SMU-114) and MOE AcRF Tier 1 grant A-8002711-00-00.

[2]Address: 15 Kent Ridge Dr, Singapore 119245. Phone: +65-6516-8119. E-mail: bizagarw@nus.edu.sg.

[3]Address: No.10 Huixin East Street, Changyang District, Beijing, China 100029. Phone: +86-010-64493592. E-mail: ydeng@uibe.edu.cn.

[4]Address: 15 Kent Ridge Dr, Singapore 119245. Phone: +65-6516-3441. E-mail: yi.fan@nus.edu.sg.

[5]Address: 90 Stamford Road, Singapore 178903. Phone: +65-8694-7689. E-mail: qi.gao.2023@phdecons.smu.edu.sg.

[6]Address: 90 Stamford Road, Singapore 178903. Phone: +65-6808-5454. E-mail: lijing@smu.edu.sg.

[7]Address: 90 Stamford Road, Singapore 178903. Phone: +65-6828-0876. E-mail: linma@smu.edu.sg.

# 1 Introduction

Housing stock in major cities ages rapidly.[1] In addition to worsening the housing supply crisis, deteriorating housing generates substantial negative externalities (Cattaneo et al., 2009; Galiani et al., 2017; Freedman and Owens, 2011; Aliprantis and Hartley, 2015; Blanco, 2023), which motivate government interventions. Housing upgrading programs represent a major form of in-kind transfer in cities worldwide, with governments committing billions to improve aging housing stock.[2] A central question for policy design is whether such in-kind provision generates positive externalities for neighboring households large enough to justify the distortions it imposes on recipients (Currie and Gahvari, 2008). The answer hinges on the magnitude and spatial reach of these externalities. In high-density urban environments, where each building is proximate to many others, upgrading externalities are potentially amplified, strengthening the case for in-kind provision. Yet whether externalities are in fact large enough to offset the distortionary cost of in-kind housing provision, and to what extent this depends on density, remain open empirical questions.

A growing body of work documents housing externalities, but most existing evidence comes from low- to moderate-density settings in the United States (Rossi-Hansberg et al., 2010; Autor et al., 2014; Hornbeck and Keniston, 2017; Fu and Gregory, 2019; Ganduri and Maturana, 2024). These studies establish that neighborhood spillovers from housing investment are positive and spatially localized, but they are estimated in settings where each property has few proximate neighbors. In high-density cities, the same per-neighbor spillovers accumulate over many more nearby properties, so the externalities at stake are potentially much larger. The existing evidence is also based almost entirely on single-family detached housing. Upgrading a multistory building is a different intervention: a single structure houses many households across multiple levels, so the nature of the interactions, and plausibly the magnitude of housing externalities, could differ substantially. Understanding how large upgrading externalities become in dense environments, and how rapidly they decay with distance, is essential for drawing policy implications.

This paper starts by estimating the magnitude and spatial reach of housing externalities in a high-density urban setting, exploiting the staggered rollout of Singapore's Main Upgrading Programme (MUP), the sole nationwide large-scale public housing upgrading initiative between 1990 and 2006. Singapore's public housing, built and managed by the Housing & Development Board (HDB), is home to about 80% of the resident population. The MUP revitalized aging HDB estates in staggered phases, upgrading 886 buildings which cover 128 (8.7%) precincts and 131,000 households and at a cost of S$3.3 billion (approximately US$2.5 billion). At the same time, Singapore's HDB estates are dense: each building has, on average, 79 neighboring buildings within 500 meters, 236 within 1,000 meters, and 426 within 1,500 meters. These features, combined with the standardized design and construction of HDB housing and the absence of overlapping large-scale policies, provides a well-suited setting for quantifying housing externalities in a dense urban environment.

[1] In the United States, for example, by 2025 about 50% of housing units were more than 45 years old, and roughly 36% were more than 55. This phenomenon is particularly acute in many high-density metropolitan areas: close to half of the housing stock in cities such as New York and Chicago was built before 1940, compared to about 12% nationwide (Source: U.S. Census Bureau American Housing Survey.)

[2] For instance, Pennsylvania's Whole-Home Repairs Program allocated over $120 million in 2022 to address habitability and safety concerns; New York City's Permanent Affordability Commitment Together (PACT) initiative has mobilized more than $5.6 billion for capital repair work; and the United Kingdom's Decent Homes Programme directed approximately £37 billion over a decade to bring public housing up to modern standards.

We exploit spatial and intertemporal variation in the policy implementation to estimate the impact of upgrading on housing values, using both two-way fixed effects (TWFE) and a staggered difference-in-differences design (Callaway and Sant'Anna, 2021). Identification rests on a parallel-trends assumption: conditional on the included fixed effects and controls, early- and later-treated buildings would, absent upgrading, have followed common counterfactual price trends. We assess this assumption with an event-study design. Because the program selectively targets aged buildings, our estimates identify the average treatment effect on the treated (ATT) for the selected pool. To measure the extent of housing externalities, we follow Miguel and Kremer (2004) and construct spillover treatment intensities by counting, for each building (whether treated or untreated), the number of treated buildings within varying distance bands of this focal building.[3] Although this approach is widely used, Borusyak and Hull (2023) and Borusyak et al. (2025) show that exogenous variation in treatment assignment does not necessarily generate exogenous variation in treatment exposure: buildings in central or densely built-up areas mechanically accumulate higher spillover intensity, so raw exposure may capture geographic characteristics rather than true externalities. To address this concern, we adopt the recentering strategy of Borusyak and Hull (2023), which controls for the *expected* treatment intensity to purge the geography-driven component of *observed* exposure.

We find positive and significant neighborhood externalities that are highly localized. After controlling for expected exposure, each additional treated building within 0–500 meters raises resale prices per square meter by 0.15%. With an average of 13 treated neighbors within this band, the implied total spillover effect is approximately 1.95%, equivalent to S$65.32 (US$49.65) per square meter. The magnitude of externalities declines with distance and approaches zero by 500–1,000 meters, consistent with strong localized demand-side effects that dominate supply-side pressure in the immediate vicinity. Upgrading also raises resale prices directly in treated buildings: unit prices increase by 1.65% (S$55.27/sqm) following the announcement and by 11.47% (S$384.18/sqm) following the completion of upgrading.

The credibility of these estimates rests on the parallel trends assumption. Event-study estimates show no evidence of differential pre-trends, and the TWFE and cohort-specific difference-in-differences estimators of Callaway and Sant'Anna (2021) yield similar magnitudes, confirming that the negative-weights concern does not substantially affect the TWFE results. A comprehensive set of robustness checks, including controls for building age, inverse probability weighting, and alternative distance bands and sample windows, confirms the stability of the own and spillover effects. Exploiting buildings that were announced for upgrading but failed the resident vote as a natural counterfactual yields consistent results.[4]

[3]A similar approach is adopted in the social network literature (Bandiera and Rasul, 2006) and the spatial spillover literature (Lu et al., 2019).

To quantify the welfare trade-off between in-kind upgrading and cash transfers, we adapt the framework of Rossi-Hansberg et al. (2010), explicitly modeling distance-decaying housing externalities and estimating key parameters via indirect inference, matching moments from our reduced-form estimates of (i) the value appreciation of treated units and (ii) the magnitude and spatial decay of externalities. The estimated model implies that externalities are quantitatively important in a dense urban setting. The upgrading program raises treated-household welfare by 2.43% without externalities and by 3.35% with externalities, reflecting both direct benefits from improved housing services and indirect benefits from neighbors' improvements; untreated neighbors gain 0.11% due to externalities alone. Comparing to an equivalent lump-sum cash transfer that allows the household to freely allocate between consumption and housing upgrading, we find that, absent externalities, cash dominates (2.53% vs. 2.43%) due to greater consumption flexibility. With externalities, cash yields smaller welfare gains for treated households (2.36% vs. 3.35%) because households underinvest in housing and free-ride on neighbors, dampening aggregate spillovers. The welfare advantage of in-kind upgrading hinges critically on population density: when neighborhood density is reduced to roughly 40% of Singapore's level (comparable to Los Angeles) or 6% (comparable to Birmingham, Alabama), spillovers weaken and the welfare ranking reverses in the latter.

To explore mechanisms underlying the spatial spillovers, we turn to administrative resident records that track individuals' registered addresses over time. Upgrading leads to a large decline in residential mobility in treated buildings, and crucially, mobility also falls in nearby untreated buildings, with effects that attenuate with distance from treated locations; resale transaction volumes mirror this spatial pattern, declining in both treated and nearby buildings. This spatial decay points to localized improvements in neighborhood conditions that raise the value of staying put not only for recipients but also for proximate non-recipients. Along with reduced turnover, treated neighborhoods experience a marked shift toward an older resident age profile, and neighboring buildings exhibit smaller but significant increases in average age as well. Together, these patterns suggest that upgrading increases place attachment, particularly among older incumbents, reshaping neighborhood composition and propagating the impact of housing upgrading beyond the directly treated buildings. This age-based retention of incumbents represents a previously underexplored channel of housing externalities, complementing the well-established physical channel—improved structures and amenities capitalizing directly into nearby prices—and the compositional channel operating through income- or education-based sorting.

[4]The government bore the majority of the upgrading cost (citizen households co-paid only 7% to 18% of the total, depending on flat type), so resident approval was nearly universal: only 1.47% of announced blocks failed the required poll. These vote-failed blocks are few but undergo the same announcement process without subsequent implementation, providing a clean within-program counterfactual.

This paper contributes to several strands of literature. First, we add to the literature measuring housing externalities and examining the mechanism through which they arise. Relative to existing estimates (Coulson and Li, 2013; Autor et al., 2014; Hornbeck and Keniston, 2017; Sandler, 2017; Fu and Gregory, 2019; Koster and Van Ommeren, 2019; Davidoff et al., 2022; Bradlow et al., 2023; Ganduri and Maturana, 2024), we use large-scale microdata and a nationwide policy to estimate the magnitude and spatial reach of externalities induced by public housing upgrading in a dense urban setting. We also contribute to work on the mechanisms by identifying a compositional channel that, to our knowledge, has not been documented previously (Harding et al., 2009; Guerrieri et al., 2013; Autor et al., 2014; Ganduri and Maturana, 2024). Housing upgrading potentially generates amenity improvements that disproportionately strengthen place attachment among older incumbent residents, reducing their residential mobility and, in turn, reshaping the age composition of both treated buildings and their immediate vicinity.

Second, this paper contributes to ongoing debates on housing subsidy design, including housing redevelopments (Jacob, 2004; Chyn, 2018; Almagro et al., 2024; Neri, 2024; Blanco and Neri, 2025), and housing vouchers (Kling et al., 2007; Baum-Snow and Marion, 2009; Ludwig et al., 2013; Pollakowski et al., 2022; Bergman et al., 2024; Chetty et al., 2026; Chyn and Daruich, 2025). The literature has documented potential drawbacks of redevelopment or relocation, including inefficiencies from landlords' strategic responses, negative supply shocks from demolition, market distortions, and welfare losses from poorly targeted transfers (Glaeser and Luttmer, 2003; Collinson and Ganong, 2018; Diamond et al., 2019; Waldinger, 2021; Majid, 2023). We study a design that preserves the existing housing stock: a public upgrading program without demolition. Because housing services generate spatial externalities, the benefits extend beyond treated households to nearby untreated neighbors. Our counterfactual analyses demonstrate that in high-density settings, direct subsidies to housing services deliver larger total welfare gains than equivalent cash transfers.

Third, more broadly, this paper speaks to the classic welfare trade-off between cash and in-kind transfers: in-kind provision can be welfare-inferior to cash because it constrains households' choice sets, but it can be welfare-superior when the targeted good is subject to market failures that cash does not correct (Currie and Gahvari, 2008). The literature has identified self-targeting (Nichols and Zeckhauser, 1982; Blackorby and Donaldson, 1988; Lieber and Lockwood, 2019), paternalism (Chorniy et al., 2025; Bandiera et al., 2023), insurance (Gadenne et al., 2024), and externalities (Cunha et al., 2019) as channels through which in-kind transfers can dominate cash. We contribute to this agenda by quantifying the externalities channel in housing, the single largest component of household consumption, with welfare estimates from a large-scale in-kind program, benchmarked against an equivalent lump-sum transfer. We also highlight that the comparison between in-kind transfers and cash critically depends on context: in housing programs, the welfare ranking hinges on population density.

Fourth, we contribute methodologically to the broad literature on causal identification of spillover effects. A growing body of work exploits random and quasi-random variation to estimate externalities (Miguel and Kremer, 2004; Bandiera and Rasul, 2006; Lu et al., 2019), but as Borusyak and Hull (2023) and Borusyak et al. (2025) highlight, exogenous variation in treatment assignment does not necessarily generate exogenous variation in treatment exposure. We implement the recentering strategy of Borusyak and Hull (2023), controlling for expected neighborhood exposure constructed from simulated counterfactual treatment timing. This approach isolates the exogenous component of realized exposure and yields a different spatial profile of externalities, underscoring the importance of accounting for nonrandom exposure in spillover analyses.

The rest of the paper is organized as follows. Section 2 provides institutional background on Singapore's public housing system and the MUP. Section 3 describes data sources and variable construction. Section 4 presents the empirical design. Section 5 reports baseline estimates, parallel trends evidence, and robustness checks. Section 6 develops the welfare model and counterfactual analyses. Section 7 investigates mechanisms. Section 8 concludes.

## 2 Institutional Background

**Public Housing in Singapore.** Public housing in Singapore is built, allocated, and managed by the HDB, which was established in 1960 with the goal of rehousing a population then living largely in slums and squatter settlements.[5] Spread across the entire island (Figure A.1), HDB flats now house about 80% of the resident population, about 90% of whom are owner-occupiers. The estates take a high-rise, high-density form: residential buildings range from about 13 to 49 storeys and, although individual layouts vary, are built from a limited set of standardized flat types, with each building housing on average 110 flats; the average building has 79 other HDB buildings within 500 meters, 236 within 1,000 meters, and 426 within 1,500 meters. Importantly, the prices we study are *not* administered. While new flats are sold by the government at subsidized, rationed prices, every transaction in our data comes from the *resale* market, in which sitting owners sell to private buyers at freely negotiated prices. Resale prices therefore reflect buyers' willingness to pay for housing and neighborhood quality, which is the object our externality estimates are meant to capture.[6]

**The Main Upgrading Programme.** Introduced in 1989, the MUP was a nationwide program to revitalize aging public housing estates, mainly targeting buildings completed up to 1980. It upgraded estates *in place*: at the building and precinct level it added or renewed shared amenities, including lifts and lift lobbies, covered walkways, drop-off porches, multi-storey car parks, and landscaped open space, while within flats it upgraded bathrooms, replaced entrance doors and grilles, and upgraded pipes and cables. Crucially for our interpretation, the program did not demolish buildings, relocate residents, or add new units: the number of flats is held fixed, so any price response mainly operates through improved amenities. The MUP was also the sole large-scale, government-sponsored upgrading program in Singapore between 1990 and 2006, which limits confounding from overlapping interventions.

[5] In 1960 only about 9% of residents lived in government flats.

[6] Eligibility is restricted to Singapore citizens and permanent residents who meet income and household criteria; owners may not hold multiple HDB units and may resell only after a Minimum Occupation Period, generally five years and up to ten or twenty for some flat types.

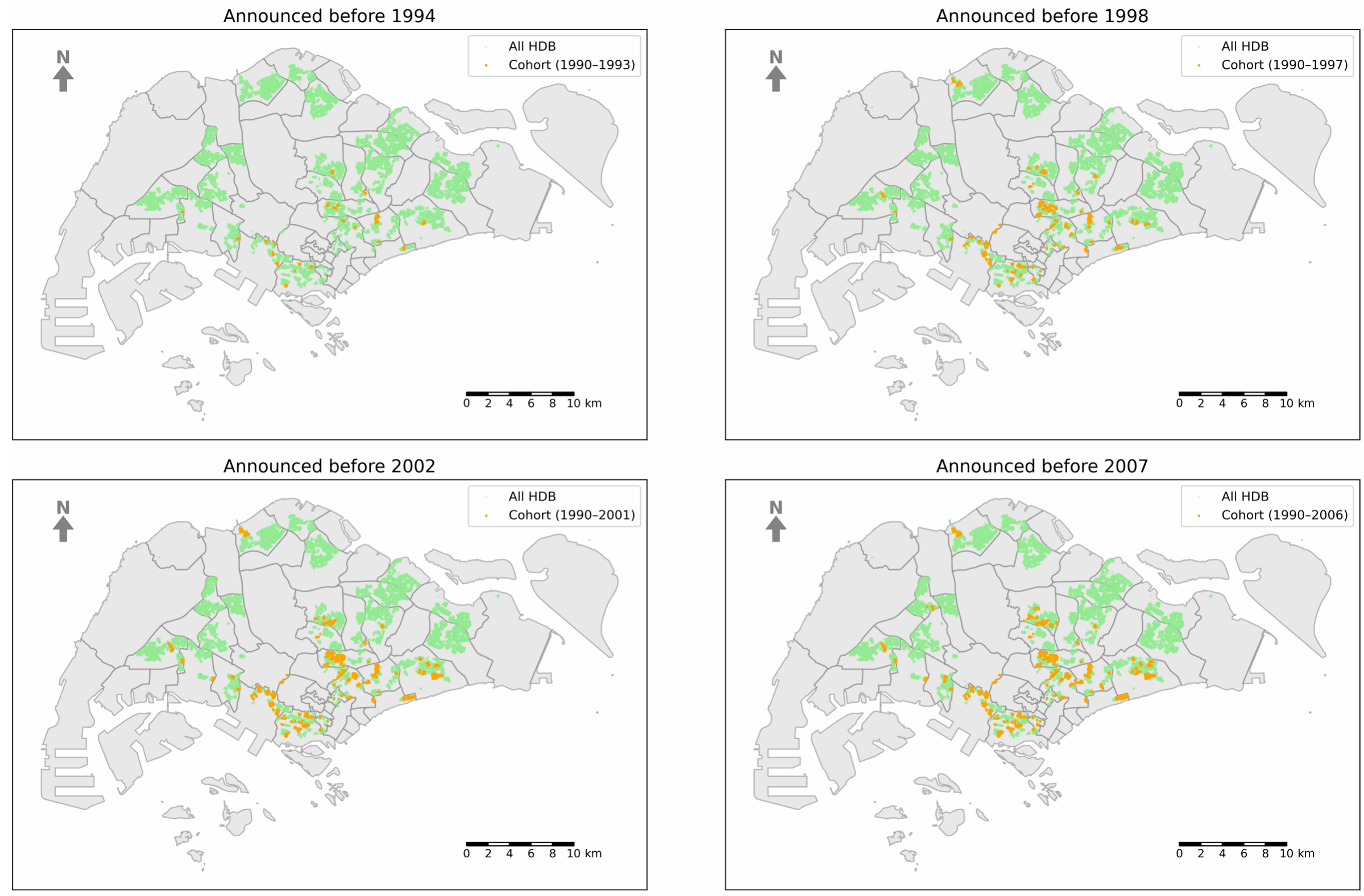

Figure 1: HDB Buildings Announced for MUP by Cohort

**Notes:** These four maps show the distribution of HDB buildings in Singapore that were announced for upgrading under the Main Upgrading Programme (MUP), grouped by cohort. Green dots represent the full universe of HDB buildings in Singapore (as of 2024), while orange dots indicate the locations of buildings selected for upgrading. In detail, the top-left map shows buildings announced between 1990 and 1993 (before 1994); the top-right map shows buildings announced between 1990 and 1997 (before 1998); the bottom-left map shows buildings announced between 1990 and 2001 (before 2002); the bottom-right map shows buildings announced between 1990 and 2006 (before 2007). Specifically, 151, 345, 280, and 110 buildings were announced for upgrading during the periods 1990–1993, 1994–1997, 1998–2001, and 2002–2006, respectively.

HDB selected precincts for upgrading primarily on the basis of building age and structural condition, with most buildings upgraded when 21 to 25 years old (Figure A.2), and proceeded only where at least 75% of resident owners approved the works in a poll.[7] Upgrading was rolled out in staggered cohorts: 151, 345, 280, and 110 buildings were announced during 1990–1993, 1994–1997, 1998–2001, and 2002–2006, respectively (Figure 1). For each building we observe two dates that anchor our design. The *announcement* date is when upgrading plans are revealed and put to the resident vote, after which the works are anticipated; the *billing* date follows immediately after upgrading construction and quality check completion, when residents are billed for their share of the cost.[8] In total, the MUP upgraded 128 precincts and benefited 131,000 households at a cost of S$3.3 billion (about US$2.5 billion). The government bore most of the cost, with citizen households co-paying only 7% to 18% of the total depending on flat type.

[7]The share of announced blocks that passed the poll is 98.53%.

Figure 2 highlights the key features for Ang Mo Kio, one of the first towns upgraded. Treated buildings are densely packed, and they enter the program in several waves spread over more than a decade. A given building is typically surrounded by many neighbors, some upgraded in earlier cohorts and some in later ones, and some remain untreated throughout. This combination of high density and staggered timing is what makes the setting well suited to studying spillovers: a building's exposure to upgraded neighbors varies both across space, with its distance to treated buildings, and over time, as nearby buildings enter the program in different years. We exploit this variation for empirical design in Section 4.

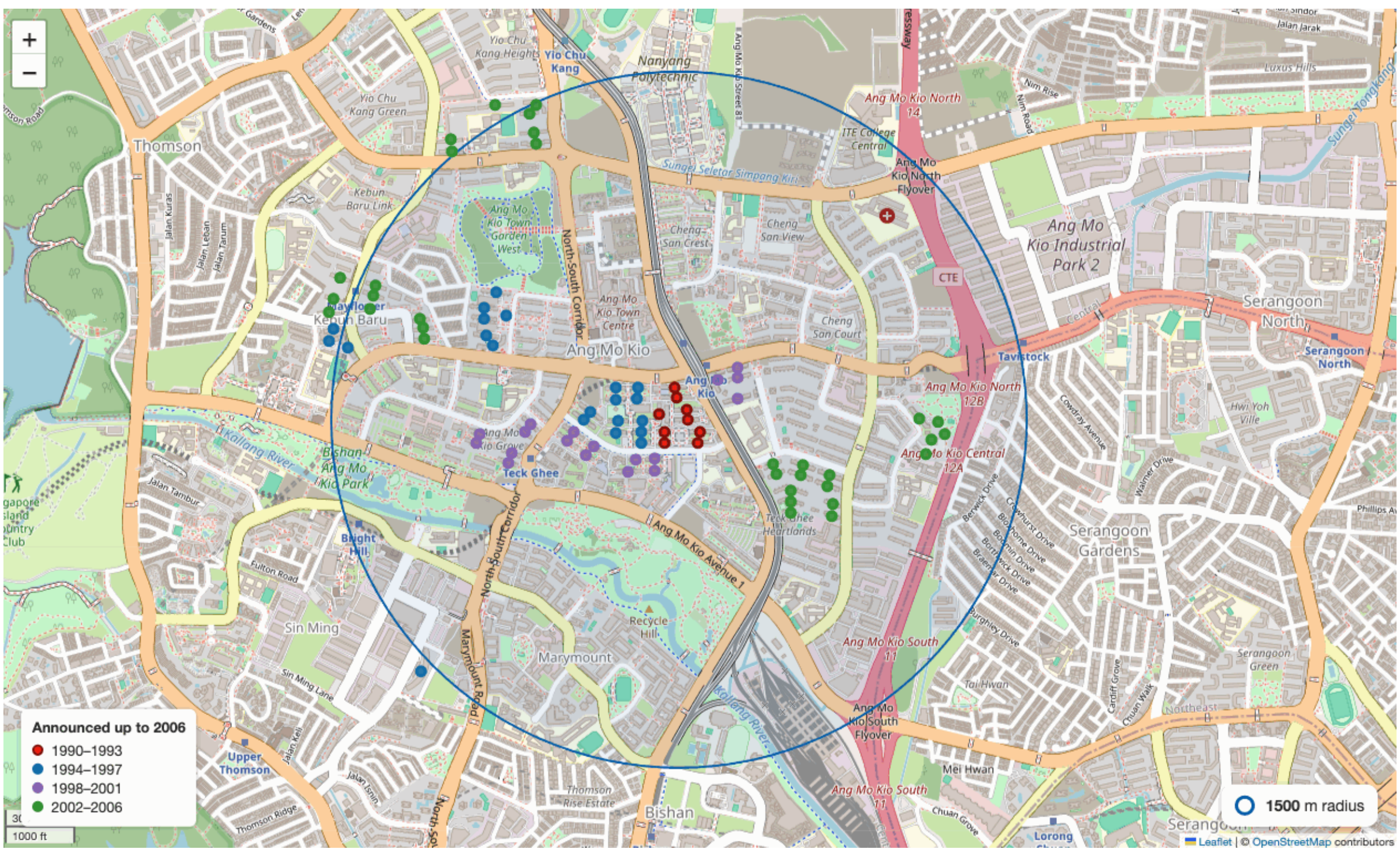

Figure 2: HDB Buildings Announced for MUP in Ang Mo Kio by Announcement Cohort

**Notes:** This map shows HDB buildings announced for upgrading under the MUP in Ang Mo Kio, grouped by announcement cohort. The circle denotes a 1,500-meter radius centered on the earliest treated cohort (1990). Red, blue, purple, and green dots indicate buildings announced in 1990–1993, 1994–1997, 1998–2001, and 2002–2006, respectively. The dense spatial clustering of treated buildings across multiple cohorts illustrates the high-density, multi-wave nature of the upgrading program in this area.

[8]The interval between the two is 6.1 years on average (Figure A.3). We separately estimate the anticipation effects from the effect of realized upgrading proxied by the billing date.

# 3 Data and Variables

## 3.1 Data

Our analysis draws on three datasets. The first is the HDB resale transaction dataset, which records the universe of HDB flat resale transactions from 1990 to 2024, totaling 942,883 sales.[9] The unit of observation is flat $i$ sold in year-month $t$. For each transaction, we observe the sale price, floor area (in square meters), storey range, flat type and model, lease-commencement year, street name, postal code, and geographic coordinates (longitude, latitude). The sale price and floor area allow us to construct our primary outcome variable, the logarithm of the resale price per square meter. In Singapore, each postal code corresponds to a unique building, a single multistory residential structure housing many individual flats; in the administrative records these buildings are designated by a *block number*, and we refer to them as buildings throughout. Each building is uniquely associated with a pair of geographic coordinates, so this one-to-one mapping between postal codes, buildings, and coordinates allows us to link datasets by postal code and to compute pairwise distances between buildings for constructing the neighborhood exposure measures described in Section 3.2.

The second dataset is HDB's administrative register of the MUP, covering the program's full operational period from 1990 to 2006. The register records all 886 upgraded buildings and, for each building, reports the block number, street name, and the exact announcement, billing, and completion dates. We map buildings to postal codes using block numbers and street names and merge the MUP register with the transaction data by postal code, which, given the one-to-one postal code–building correspondence, yields an exact match at the building level. Full details on MUP implementation schedule are provided in Online Appendix Table A.1.

The third dataset is a confidential administrative dataset of 2,171,383 Singaporean residents aged 20 or older, drawn from government registers linked to individuals' official identification records.[10] For each individual, we observe gender, age, residential address (identified by postal code), and housing type at discrete waves from 1996 to 2018. Approximately 99.4% of the observations in our analysis fall in the 1996, 1998, 2000, 2005, and 2011 waves, which cover a substantial portion of the MUP's operational period and allow us to track the impact of the program over time at the individual level. Because addresses are recorded repeatedly, we can measure residential mobility (whether an individual changes building between consecutive waves) and examine how the demographic composition of residents evolves at the building level following the MUP. We use these outcomes in Section 7 to investigate the behavioral mechanisms underlying the price effects.

[9] Resale transactions are secondary-market sales between existing owners and buyers, distinct from the initial allocation of new flats by the government; resale prices are therefore market-determined.

[10] Based on the Department of Statistics Singapore's mid-year 2019 estimates, Singapore's total population was 5,703,600, including 4,026,200 Singapore citizens or permanent residents. The population aged 20 and above was 3,213,000.

## 3.2 Variables

**Direct Treatment Status.** The MUP proceeds through two distinct phases for each treated building: announcement and billing. We construct two binary indicators to capture the direct treatment status of building $b(i)$, which contains HDB flat $i$, in year-month $t$. The indicator $\mathbb{1}(announce)_{b(i),t}$ equals one if the transaction occurs after the building's official announcement date but before the billing date, and zero otherwise. The indicator $\mathbb{1}(billing)_{b(i),t}$equals one if the transaction occurs on or after the billing date, and zero otherwise. These variables identify the own effect of upgrading on treated buildings across the two policy phases, allowing us to distinguish the anticipation response following announcement from the effect of actual implementation.

**Neighborhood Exposure.** To measure the spatial reach of upgrading spillovers, we construct distance-banded measures of neighborhood exposure. Using each building's geographic coordinates, we compute pairwise Euclidean distances $d_{jk}$ between all HDB buildings $j$ and $k$, and for each building (both treated and untreated), we count the number of neighboring buildings that have entered each upgrading phase within distance bands $r \in \{0\text{–}500\,\text{m},\ 500\text{–}1000\,\text{m},\ 1000\text{–}1500\,\text{m}\}$ around this focal building:

$$Ann_neighbor^{r}_{b(i),t} = \sum_{k \neq b(i)} \mathbb{1}\{d_{b(i)k} \in r\} \cdot \mathbb{1}(announce)_{k,t}$$

$$Bill_neighbor^{r}_{b(i),t} = \sum_{k \neq b(i)} \mathbb{1}\{d_{b(i)k} \in r\} \cdot \mathbb{1}(billing)_{k,t}$$

where $\mathbb{1}\{d_{b(i)k} \in r\}$ equals one if the distance between buildings b(i) and k falls within band r.

Thus $Ann_neighbor^{r}_{b(i),t}$ and $Bill_neighbor^{r}_{b(i),t}$ count the number of neighboring buildings within band r that have entered each upgrading phase (between announcement and billing/completion or after billing/completion) by date t. We further define the cumulative count of post-announcement treated neighbors, $Neighbor^{r}_{b(i),t} \equiv Ann_neighbor^{r}_{b(i),t} + Bill_neighbor^{r}_{b(i),t}$ , which records all neighbors within band r whose upgrading has been announced by date t, regardless of billing status.

**Expected Neighborhood Exposure.** Realized neighborhood exposure may reflect underlying geographic patterns rather than causal spillovers: buildings in central or densely built-up areas mechanically have more neighbors and therefore tend to exhibit higher exposure under even random rollout schedule. To isolate the exogenous variation in exposure, we construct an *expected neighborhood exposure* following the recentering strategy of Borusyak and Hull (2023). Specifically, we hold the set of 886 upgraded buildings fixed and retain the realized number of buildings announced in each month, but randomly reassign which buildings receive which announcement dates across 5,000 simulations. For each simulation, we recompute the number of upgraded neighbors within each distance band and average these simulated counts to obtain the expected exposure. This measure captures the neighborhood exposure a building would receive purely due to its geographic position and local density under random assignment, holding the aggregate rollout schedule fixed. Including expected exposure as a control in the regression absorbs the predictable, geography-driven component of realized exposure, so that identifying variation comes from deviations of actual exposure from its expected level. Details of the simulation algorithm are provided in Online Appendix Section D.

**Summary Statistics.** Our estimation sample comprises 531,836 HDB resale transactions in treated buildings and in buildings with at least one treated neighbor within 1,500 meters, the maximum distance over which we measure exposure. Online Appendix Table C.1 reports summary statistics for this sample. The average resale price per square meter in our sample is S$3,349.4 (US$2,642.4), with a mean floor area of 89.9 square meters, both broadly representative of the national housing stock. Approximately 5.7% of transactions occur in the post-announcement, pre-billing window, and 12.2% in the post-billing period.

# 4 Empirical Design

Our goal is to estimate both the direct effect of upgrading on treated buildings and its spillover effects on nearby buildings. Although upgrading is expected to raise prices within treated buildings, its effects on nearby neighborhoods are theoretically ambiguous. On the demand side, upgrading improves local amenities and may attract higher-income households, increasing nearby willingness to pay. On the supply side, upgrading expands the local stock of housing services, which could place downward pressure on prices. The relative strength of these channels is also likely to vary with distance: demand effects are expected to dominate in the immediate vicinity, whereas supply effects may extend over a broader area. Moreover, the MUP proceeds in distinct phases (announcement and billing), which may generate different price responses as information about upgrading is gradually revealed and capitalized. These considerations motivate us to begin with a flexible specification that separately identifies own effects by upgrading phase and neighborhood spillovers by both phase and distance band.

We begin with a TWFE specification that lets the own effect and neighborhood spillovers vary freely across upgrading phases and distance bands:

$$
\begin{aligned}
\ln\left(\text{price_psm}_{i,t}\right) = {} & \beta_1 \cdot \mathbb{1}(\text{announce})_{b(i),t} + \beta_2 \cdot \mathbb{1}(\text{billing})_{b(i),t} \\
& + \beta_3 \cdot \text{Ann_neighbor}^{0-500}_{b(i),t} + \beta_4 \cdot \text{Ann_neighbor}^{500-1000}_{b(i),t} + \beta_5 \cdot \text{Ann_neighbor}^{1000-1500}_{b(i),t} \\
& + \beta_6 \cdot \text{Bill_neighbor}^{0-500}_{b(i),t} + \beta_7 \cdot \text{Bill_neighbor}^{500-1000}_{b(i),t} + \beta_8 \cdot \text{Bill_neighbor}^{1000-1500}_{b(i),t} \\
& + X_{i,t}{}' \cdot \beta_9 + \alpha_{b(i)} + \alpha_{ym(t)} + \alpha_{s(i)} \cdot y(t) + \varepsilon_{i,t}
\end{aligned}
$$

where i represents individual HDB flat, b(i) denotes the building containing flat i, and t is the transaction period at the year-month level. The dependent variable, $\ln\left(\text{price_psm}_{i,t}\right)$, is the log resale price per square meter of flat i transacted in period t. As detailed in Section 3.2, the indicators $\mathbb{1}(announce)_{b(i),t}$ and $\mathbb{1}(\text{billing})_{b(i),t}$ capture the upgrading phases of the treated buildings. The varibles the number of neighboring treated buildings (excluding the building itself) within distance band r that have entered the corresponding phase, with r denoting the bands of 0–500m, 500–1,000m, and 1,000–1,500m. We control for flat characteristic $X_{i,t}$, including flat size (in square meters) and fixed eff ects for flat type, storey range, and flat model. We also include building fixed effects $\alpha_{b(i)}$, year-month fixed effects $\alpha_{ym(t)}$, and street-specific year trends $\alpha_{s(i)} \cdot y(t)$, where $s(i)$ denotes the street, that is, the road-name grouping recorded in the transaction data that spans several adjacent buildings.

**Controls and Identification.** The inclusion of a rich set of controls ensures that $\beta_1 - \beta_8$ capture differential price movements within local housing markets rather than the compositional differences across areas. Conditional on these controls, the treatment-phase indicators identify own-building effects from within-building timing variation, while the exposure terms identify externalities from variation in the number of treated neighbors within each distance band. The key identifying assumption is that, conditional on the fixed effects and controls, early- and later-treated buildings would, absent upgrading, have followed common counterfactual price trends. We assess this assumption using an event-study design in Section 5 that tests for differential pre-trends before the announcement date. Because the program selectively targets aged buildings, the estimated own and neighboring effects identify the ATT for the program's selected pool, not an average effect that can be extrapolated to arbitrary buildings.

**Staggered Treatment Timing.** A well-known concern with TWFE estimation under staggered adoption is that already-treated units may serve as implicit controls for later-treated ones, producing non-convex or negative weights on cohort-time effect (De Chaisemartin and d'Haultfoeuille, 2020; Goodman-Bacon, 2021). Bias arises in particular when later-treated cohorts are compared to earlier-treated "controls" that are themselves contaminated by treatment dynamics. Moreover, as Goodman-Bacon (2021) shows, the TWFE estimator is a weighted average of all possible pairwise DiD estimates, where the weights are "variation hungry" and do not generally correspond to the standard weights used for the ATT. To verify that this does not distort our estimates, we additionally implement the cohort-specific difference-in-differences (CSDID) estimator of Callaway and Sant'Anna (2021), which compares each treated cohort only to not-yet-treated units and aggregates group-time effects using a well-defined convex weighting scheme. We apply CSDID to the own treatment effects; the neighboring exposure variables, which are continuous and time-varying, fall outside the scope of the Callaway and Sant'Anna (2021) framework. As we show in Section 5, the TWFE and CSDID estimates yield very similar magnitudes, suggesting that the problematic comparisons do not carry high weights in generating the TWFE estimates and that the negative-weights concern does not substantially affect these estimates.

**Nonrandom Neighborhood Exposure.** Despite the rich set of controls, the neighborhood exposure measures may still be subject to omitted-variable bias: the extent of spillover treatment a unit experiences may be systematically correlated with its position in the spatial network, even when the treatment itself is as good as randomly assigned (Borusyak and Hull, 2023; Borusyak et al., 2025). In our setting, buildings in central or densely built-up areas mechanically have more neighbors and tend to accumulate higher exposure under even random rollout schedule, so raw exposure may capture unobserved geographic characteristics (e.g., centrality, accessibility) that evolve with the program rather than true spillovers. To address this concern, we include the expected neighborhood exposure constructed in Section 3.2 as an additional control. This measure absorbs the predictable, geography-driven component of realized exposure, so that identification of the neighboring effects comes from deviations in the timing and spatial placement of actual upgrades rather than from structural differences in neighborhood density or location fundamentals.

**Baseline Specification.** Following the above discussions, we further revise Equation (1) by adding expected neighborhood exposure controls. Additionally, because the estimated neighborhood effects are similar across the announcement and billing phases (Online Appendix Table C.2), we collapses the phase-specific neighbor counts into a single cumulative count of post-announcement treated neighbors to form our preferred baseline specification:

$$\begin{aligned}\ln\left(\text{price_psm}_{i,t}\right) &= \gamma_1 \cdot \mathbb{1}(\text{announce})_{b(i),t} + \gamma_2 \cdot \mathbb{1}(\text{billing})_{b(i),t} \\ &+ \sum_r \theta_r \cdot \text{Neighbor}^r_{b(i),t} + \sum_r \lambda_r \cdot \overline{\text{Neighbor}}^r_{b(i),t} \\ &+ X_{i,t}{}' \cdot \delta + \alpha_{b(i)} + \alpha_{ym(t)} + \alpha_{s(i)} \cdot y(t) + \varepsilon_{i,t}\end{aligned}$$

where $\text{Neighbor}^r_{b(i),t}$ is the cumulative number of treated neighbors within band r whose upgrading has been announced by t (equivalently, $Ann_neighbor^r_{b(i),t} + Bill_neighbor^r_{b(i),t}$from Section 3.2), and $\overline{\text{Neighbor}}^r_{b(i),t}$is the corresponding expected exposure from Section 3.2; rindexes the same three distance bands, and the controls $X_{i,t}$ are as in Equation (1).

**Inference.** We cluster standard errors at the building level. Because our data comprise the universe of HDB resale transactions rather than a random sample, the uncertainty relevant for our causal estimands is design-based, arising from the assignment of upgrading across buildings and over time, rather than sampling-based (Abadie et al., 2020); building-level clustering permits arbitrary correlation among the repeated transactions of a given building across years and across its announcement and billing phases, and matches the level at which the own-treatment indicators are defined (Abadie et al., 2023). A remaining concern is that resale prices may also be spatially correlated across nearby buildings, which would motivate clustering at a broader geographic level. Much of this dependence is absorbed by the mean structure; for the spillover terms, recentering on expected exposure further purges the predictable, geography-driven component of cross-building dependence. We nonetheless verify in Section 5.3 that our conclusions are unchanged when standard errors are clustered at a broader, precinct-scale residential-cluster level that allows arbitrary correlation among nearby buildings (Online Appendix Table C.10).

**Additional Identification Concerns.** Several additional threats to identification merit attention. First, treatment selection is nonrandom if the program systematically targets buildings with particular characteristics (e.g., building age) that are themselves correlated with price trends. Second, unobserved time-varying shocks may coincide with the upgrading timeline, generating spurious treatment effects. Third, anticipation effects could arise if market participants expect nearby buildings to be upgraded in the future, confounding the estimated spillovers with expectation premia. We address each of these concerns through a comprehensive set of robustness checks in Section 5.3, including controls for building age, inverse probability weighting, heterogeneity analysis by building age, alternative distance bands, alternative sample windows, and a comparison of vote-passed versus vote-failed buildings.

# 5 Results

## 5.1 Baseline Estimates

We begin with the general specification in Equation (1), estimated on HDB resale transactions from 1990 to 2024; Online Appendix Table C.2 reports the results, distinguishing the announcement and billing phases for both the own and neighboring effects. The estimation sample comprises treated buildings together with buildings that have at least one treated neighbor within the widest distance band included in a given column: Column (1) includes only treated buildings (own effects alone), and Columns (2)–(4) progressively add the 0–500m, 500–1,000m, and 1,000–1,500m neighbor bands and the buildings exposed within them, so the number of observations rises from 132,759 to 531,836.

The MUP has a direct and positive effect on resale prices of treated buildings, with significant coefficients in both the post-announcement, pre-billing and the post-billing periods. Neighborhood exposure also raises prices, indicating positive housing externalities. The neighboring effects are similar in magnitude across the two phases, suggesting that the market capitalizes expected amenity gains at announcement. Once upgrading is announced, nearby buildings price in essentially the same stream of improvements regardless of whether billing has commenced, so the relevant state variable for neighborhood exposure is the cumulative stock of announced neighbors.

Motivated by the similarity of neighboring effects across phases, we collapse the two phase-specific neighbor counts into a single cumulative count of post-announcement treated neighbors. Columns (1)–(3) of Table 1 report this specification before correcting for nonrandom exposure. In Column (3), resale prices of treated flats increase by 1.16% in the post-announcement, pre-billing period and by 12.44% after billing. For neighborhood exposure, one additional treated building within 0–500m, 500–1,000m, and 1,000–1,500m raises unit prices by 0.55%, 0.19%, and 0.05%, respectively.

Table 1: The Impact of Main Upgrading on Housing Prices

| | Without Expected-Exposure Control | | | With Expected-Exposure Control | | |
|---|---|---|---|---|---|---|
| | (1) | (2) | (3) | (4) | (5) | (6) |
| Own Effect – After Announcement and Before Billing | 0.0088 | 0.0107* | 0.0116* | 0.0139** | 0.0172*** | 0.0165*** |
| | (0.0062) | (0.0062) | (0.0063) | (0.0062) | (0.0062) | (0.0062) |
| Own Effect – After Billing | 0.1168*** | 0.1196*** | 0.1244*** | 0.1116*** | 0.1130*** | 0.1147*** |
| | (0.0101) | (0.0101) | (0.0102) | (0.0103) | (0.0103) | (0.0103) |
| Neighboring Effect (0-500m) – After Announcement | 0.0048*** | 0.0051*** | 0.0055*** | 0.0019*** | 0.0014*** | 0.0015*** |
| | (0.0004) | (0.0004) | (0.0004) | (0.0003) | (0.0003) | (0.0003) |
| Neighboring Effect (500-1000m) – After Announcement | | 0.0016*** | 0.0019*** | | 0.0000 | 0.0000 |
| | | (0.0003) | (0.0003) | | (0.0002) | (0.0002) |
| Neighboring Effect (1000-1500m) – After Announcement | | | 0.0005** | | | -0.0010*** |
| | | | (0.0002) | | | (0.0002) |
| $N$ | 334878 | 437488 | 531836 | 334878 | 437488 | 531836 |
| adj. $R^2$ | 0.961 | 0.958 | 0.956 | 0.961 | 0.959 | 0.956 |
| Year x Month FE | Yes | Yes | Yes | Yes | Yes | Yes |
| Street x Year Trend | Yes | Yes | Yes | Yes | Yes | Yes |
| Building FE | Yes | Yes | Yes | Yes | Yes | Yes |
| Flat FE | Yes | Yes | Yes | Yes | Yes | Yes |

*Notes:* This table reports estimates of the effects of the MUP on HDB resale prices. In each column, the sample comprises treated buildings and those with at least one treated neighbor within the widest distance band included in that column; observation counts therefore increase as wider bands are added, and Columns (4)–(6) use the same samples as Columns (1)–(3). The dependent variable, ln($price\ psm_{i,t}$), is the log resale price per square meter of flat $i$ transacted in period $t$. "Own Effect" refers to binary indicators that equal one if the transaction date is on or after the start of the corresponding policy phase and zero otherwise: "After Announcement and Before Billing" denotes the window between the announcement and billing dates, and "After Billing" the period after the billing date. "Neighboring Effect" is the cumulative number of post-announcement treated neighbors within the indicated distance band (0–500m, 500–1,000m, or 1,000–1,500m). Columns (1)–(3) report estimates without controlling for expected exposure; Columns (4)–(6) additionally control for the expected neighborhood exposure within each band, constructed by simulating counterfactual treatment timing that preserves the schedule of policy implementation while randomly reassigning which buildings are treated at what time and averaging the simulated exposures over draws. Within each panel, the columns progressively add the 0–500m, 500–1,000m, and 1,000–1,500m bands. All specifications control for flat size and include fixed effects for flat type, storey range, flat model, building, and year-month, as well as street-specific year trends. Standard errors clustered at the building level are reported in parentheses. * p < 0.1, ** p < 0.05, *** p < 0.01.

As discussed in Section 4, raw neighborhood exposure may be confounded by geography-driven variation. Our preferred baseline, Equation (2), therefore adds the expected neighborhood exposure as a control; Columns (4)–(6) of Table 1 report these estimates. The own effects remain stable after this correction: the post-billing own effect is 11.47% (Column (6)). The neighboring effects decline, however. Only treated buildings within 500 meters retain a significant positive effect, with each additional treated neighbor raising unit prices by 0.15%. With an average of 13 treated neighbors within 500 meters, the implied aggregate spillover effect within this range is approximately 1.95%. Effects at 500–1,000m are negligible, and those at 1,000–1,500m are negative. The results reveal a clear spatial profile of upgrading externalities.

## 5.2 Parallel Trends

The credibility of the difference-in-differences estimates rests on the parallel trends assumption: conditional on the fixed effects and controls described in Section 4, the timing of upgrading must be uncorrelated with unobserved determinants of price changes. We assess this assumption using an event-study design centered on the MUP announcement date, with the year immediately preceding announcement ($k = -1$) as the reference period.

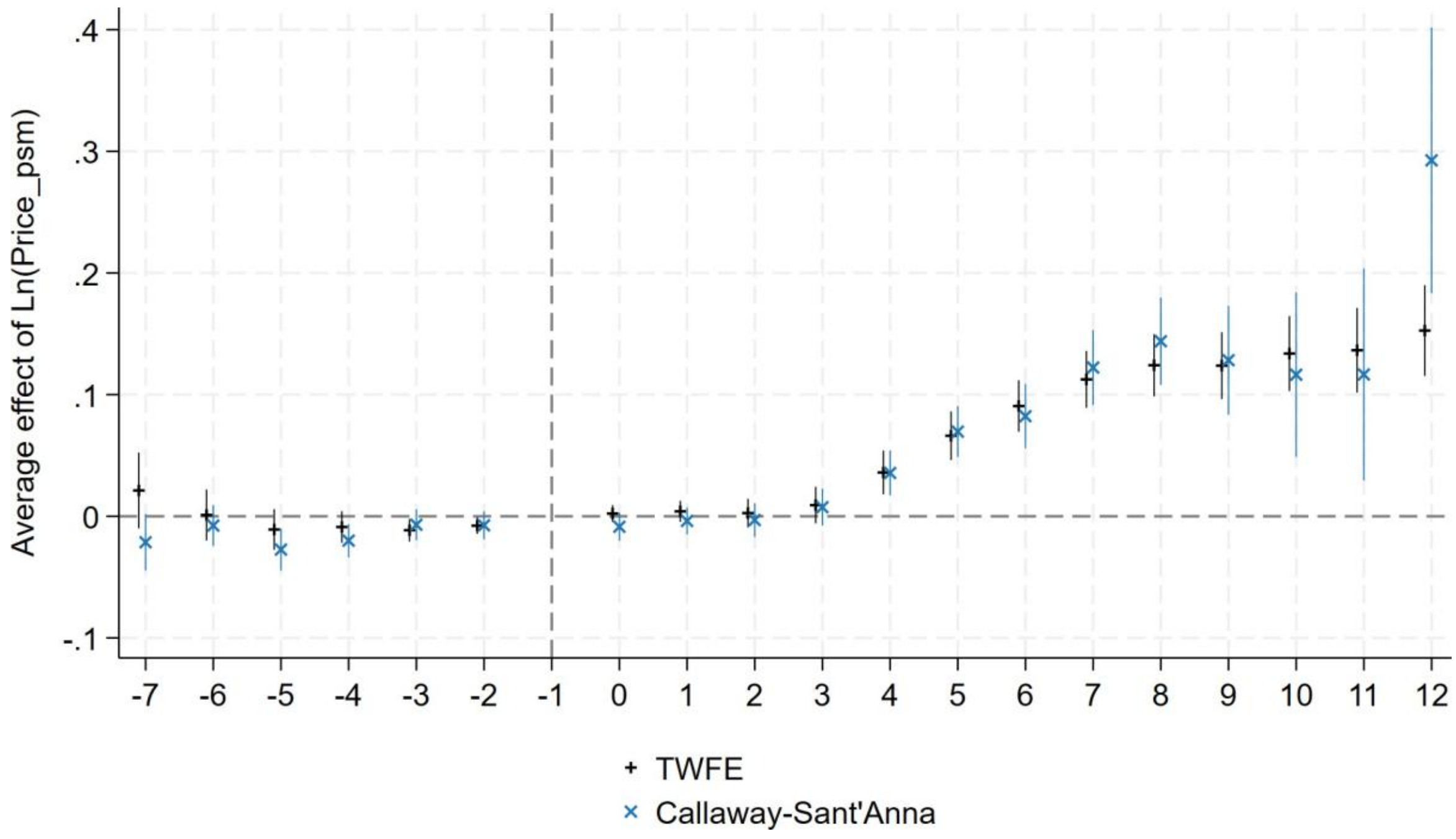

Figure 3: Event-Study Estimates of MUP Effects on HDB Resale Prices

**Notes:** This figure plots the dynamic effects of the MUP on HDB resale prices, using the year immediately preceding the announcement ($k = -1$) as base period. The event-study sample is restricted to treated buildings. The dependent variable, $\ln(price_psm_{i,t})$, is the log resale price per square meter of flat $i$ transacted in period $t$. Coefficients are estimated from a specification with leads and lags of treatment, controlling for flat size and including fixed effects for flat type, storey range, flat model, building, and year-month, as well as street-by-year trends. Markers denote point estimates, and vertical bars denote 99% confidence intervals based on standard errors clustered at the building level. In addition to TWFE estimates, the figure reports estimates from CSDID method (Callaway and Sant'Anna, 2021; Sant'Anna and Zhao, 2020), which accommodates variation in treatment timing across units and allows for heterogeneous treatment effects.

Figure 3 reports both the TWFE event-study estimates and the CSDID estimates of Callaway and Sant'Anna (2021). The lead coefficients are close to zero, providing no evidence of differential pre-trends. The post-announcement coefficients are positive and statistically significant, indicating a clear price response to upgrading. The two estimators yield similar magnitudes, confirming that the negative-weights concern under staggered treatment timing does not substantially affect our estimates. A further concern is that the not-yet-treated buildings serving as comparison units may themselves already be exposed to nearby upgrading, which would distort the estimated own effect. Re-estimating the event study on a clean-control sample that excludes comparison observations with at least one treated neighbor within 500 meters leaves the pattern essentially unchanged (Online Appendix Figure B.1).

We also implement the HonestDiD sensitivity analysis of Rambachan and Roth (2023) for the direct price effect. Rather than imposing exact parallel trends, this procedure allows post-treatment deviations from parallel trends to be bounded relative to the largest pre-treatment deviation. Online Appendix Figure B.2 shows that the average post-announcement own-building effect remains positive under small relaxations of parallel trends, with robust confidence intervals excluding zero until the relative-magnitude bound reaches roughly 0.30. The direct effect is therefore robust to modest, though not arbitrary, violations of parallel trends.

**Price Dynamics.** The event-study pattern also reveals that prices do not capitalize the full upgrading premium at announcement. Consistent with Table 1, the own effect between announcement and billing is small relative to the post-billing effect. The dynamic response is initially modest and strengthens substantially about four years later, roughly when the earliest-billed buildings enter the billing phase, and continues to build as more cohorts are billed. Although housing markets are forward-looking, several frictions together with construction externalities may delay full adjustment: information about the scope, timing, and quality of works arrives gradually, and upgrades within a building are often staged, so the ultimate features and completion dates are not fully resolved at announcement. These factors attenuate the initial response and generate a larger price adjustment once billing credibly commits the project.

### 5.3 Heterogeneity and Robustness

**Heterogeneity by Density.** Given the emphasis on the role of density, we first ask whether the estimated externalities themselves vary with local density. We split the sample at the median number of HDB buildings within 1,500 meters of a transacted building (339 buildings) and re-estimate the baseline specification, separately for below- and above-median-density neighborhoods. Online Appendix Table C.3 reports the results. The near-distance spillover is markedly stronger where density is higher: in the full specification (Columns (3) and (6)), each additional treated building within 500 meters raises unit prices by 0.24% in above-median-density neighborhoods, nearly three times the 0.09% estimated in below-median-density neighborhoods. The own post-billing effect is likewise larger in denser areas (13.4% versus 9.0%). Beyond 500 meters the realized spillovers are small and, in sparser neighborhoods, slightly negative, consistent with dispersion forces dominating where treated buildings are more isolated. This density gradient in the estimated externalities implies that the spillover benefits that make in-kind upgrading attractive are likely concentrated in the dense environments.

**Treatment Selection.** Upgrade assignment may not be random. Online Appendix Figure A.2 shows that upgrading is concentrated in a narrow age range, with most buildings upgraded at 21–25 years, indicating that selection is closely tied to building age. Although we do not claim to estimate the average treatment effect, we pursue following strategies to investigate potential implications: (i) controlling directly for building age at the time of transaction (Online Appendix Table C.4); (ii) implementing inverse probability weighting (IPW) following Abadie (2005), with treatment propensities estimated from a range of observed building characteristics, including building age (Online Appendix Tables C.5 and C.6). Both the own and neighboring effects remain stable across all three specifications. Further details are provided in Online Appendix Section E.1.

**Upgrading Expectations.** A potential concern is that our estimates of spatial spillovers may capture an "expectation premium" (pre-existing market expectations of future upgrading rather than the causal effects of MUP itself). To probe this channel, we split the sample by building age at the time of transaction (Age $\geq 14$ versus Age $< 14$) and re-estimate the baseline specification within each subsample. Online Appendix Table C.7 reports the results. For older buildings (Column (1)), the neighboring effects are small and statistically indistinguishable from zero, indicating that nearby upgrading does not induce a speculative price run-up for older units. By contrast, newer buildings (Column (2)) exhibit positive and precisely estimated spillover effects. Because newer units are less likely to be perceived as imminent candidates for upgrading, this pattern is difficult to reconcile with an expectation-based channel and is instead consistent with spillovers operating through realized improvements in local housing services.

**Alternative Distance Bands.** To verify that our baseline estimates are not driven by the choice of a 500m exposure band, we split the 0–500m band into two narrower bands (0–200m and 200–500m) and re-estimate the specification. Online Appendix Tables C.8 and C.9, which report the separate-phase and cumulative specifications respectively, show that the estimated coefficients associated with the two nearest bands are quantitatively similar, justifying the grouping in the baseline specification. We also find that the estimated effects remain stable in sign and magnitude, with a clear distance gradient, confirming that the spillovers are robust to finer spatial resolution.

**Alternative Clustering Level.** As discussed in Section 4, we also assess sensitivity to spatial correlation in resale prices by clustering at a broader spatial level. We group nearby buildings into compact, precinct-scale residential clusters constructed from the pairwise distance matrix, with construction and validation detailed in Online Appendix Section E.2, and re-estimate the baseline clustering standard errors at this level. As shown in Online Appendix Table C.10, the own post-billing effect and the near-field (0–500m) spillover remain statistically significant, with only modestly larger standard errors.

**Alternative Sample Window.** The event-study pattern (Figure 3) shows that treatment effects rise in the early years and largely plateau by year 8. We restrict the estimation sample to an eight-year window following the announcement date to reduce the scope for slow-moving secular trends or later confounding shocks to influence the estimates.[11] Online Appendix Table C.11 shows that the magnitudes and significance levels remain stable relative to the baseline.

**Vote-Failed Buildings.** Buildings selected for upgrading may be located in areas with stronger underlying growth potential, generating faster price appreciation even absent upgrading. To examine this concern, we exploit buildings that were announced for MUP but ultimately failed the resident vote and therefore were not upgraded. These vote-failed buildings provide a natural counterfactual: they undergo the same announcement process but do not receive subsequent implementation. We interact $\mathbb{1}(\text{pass})_{b(i),t}$ (whether the vote passes) with $\mathbb{1}(\text{announce})_{b(i),t}$ and $\mathbb{1}(\text{billing})_{b(i),t}$ within the pool of announced buildings and using the same controls and fixed effects as our main specification.[12] Online Appendix Table
C.12 shows that vote-failed buildings experience a decline in resale prices after announcement. If post-announcement appreciation were driven by favorable location fundamentals, prices in vote-failed buildings should continue to rise even without implementation. Instead, the negative response suggests that the announcement and subsequent vote failure jointly convey adverse information. The estimated effects for vote-passed buildings remain close to the baseline estimates in Online Appendix Table C.2 (Column (1)), indicating that our findings are not driven by selective placement of announced buildings in high-growth locations.

# 6 Welfare Implications

The presence of housing externalities points to an important trade-off in evaluating housing upgrading policies. As an in-kind transfer, upgrading improves unit quality and increases housing services, but may yield lower welfare than an equivalent lump-sum cash transfer due to distortions of households' allocation choices. When housing services generate positive local externalities, however, nearby households also benefit from the subsidy. In sufficiently dense neighborhoods, these spillover gains can outweigh the distortion cost of in-kind provision. The net welfare effect of upgrading relative to cash transfers therefore depends on the magnitude of externalities accruing to non-recipients relative to the distortions borne by recipients. To assess this trade-off quantitatively, we follow Rossi-Hansberg et al. (2010) and estimate a simple model using our reduced-form estimates.

[11] Because our sample spans a long period, estimates based on very long post-treatment windows may be more sensitive to confounding forces that are difficult to absorb even with rich fixed effects. Concentrating on the eight-year horizon targets the period in which price capitalization and spillovers are most economically relevant.

[12] More technical details are provided in Online Appendix Section E.3.

## 6.1 Model

We build a simple closed-city model tailored to Singapore's institutional setting. The city comprises a finite set of discrete locations $i \in N = \{1, \ldots, I\}$; agents are identical, supply one unit of labor inelastically, and earn wage income $w > 0$, which is allocated between consumption $c(i)$ and housing expenditure $h(i)$. Preferences are Cobb–Douglas in consumption and housing services, $u(i) = c(i)^{\alpha}\tilde{h}(i)^{1-\alpha}$ , where effective housing services are given by

$$\tilde{h}(i) = h(i) + \delta_1 \sum_{k \neq i} e^{-\delta_0 d_{ik}} h(k),$$

with $\delta_1 > 0$ capturing the intensity of externalities and $\delta_0 > 0$ governing the spatial decay rate. Taking neighbors' expenditures as given, each household chooses $(c(i), h(i))$ subject to $c(i) + h(i) = w$, yielding a neighborhood equilibrium profile $\{h^*(i)\}_{i=1}^{I}$ satisfying the best-response system implied by the externality structure (full model derivations are provided in Online Appendix Section F; existence and uniqueness are established in Online Appendix Section F.6).

We interpret the MUP as an in-kind subsidy that increases housing services by $\sigma > 0$ in the upgraded locations $A \subseteq N$. The policy is modeled as a direct addition to housing service consumption in treated locations, so untreated locations are affected only through the externality term. This structure delivers two implications: (i) treated households more exposed to other treated buildings experience larger welfare changes, and (ii) spillovers raise welfare for untreated locations absent direct transfers. We estimate $(\delta_0, \delta_1, \sigma)$ by indirect inference, targeting the own treatment effect on housing values and the magnitude and spatial decay of neighborhood externalities from Section 5. Estimation details can be found in Online Appendix Section F.4.

Although Section 5 shows that the estimated coefficients vary with neighborhood density, we calibrate the model parameters $(\delta_0, \delta_1, \sigma)$ to our baseline estimates, which pool all buildings across Singapore. Singapore lies near the high end of the global density distribution, and the lower-density environments we examine in the counterfactuals below fall well into the left tail, so the pooled estimates best characterize the high-density setting that anchors our analysis. One could instead extrapolate the estimated density gradient in the coefficients to infer how the magnitude of externalities changes across the density distribution beyond the range of our setting, but such an exercise would rely on strong functional-form assumptions, and we do not pursue it here. Relatedly, because the reduced-form estimates identify the ATT for the program's selected pool of aged buildings, the welfare numbers below should be read as policy-relevant for upgrading targeted at similar buildings.

## 6.2 Welfare Counterfactuals

Before presenting the aggregate welfare results, we illustrate the model-implied welfare changes using Ang Mo Kio again as an example. Welfare changes are computed relative to the pre-subsidy baseline using the estimated model described above.

Figure 4 maps the model-implied utility gains for upgraded buildings and surrounding buildings within a 1,500-meter radius; the spatial layout of the town's upgrading cohorts is shown earlier as in Figure 2 of Section 2. Treated buildings exhibit the largest utility increases, reflecting both the direct treatment effect and the spillovers from nearby treated neighbors. Untreated neighboring buildings also experience positive gains that decline with distance from treated sites, consistent with spatial spillovers in housing upgrading.

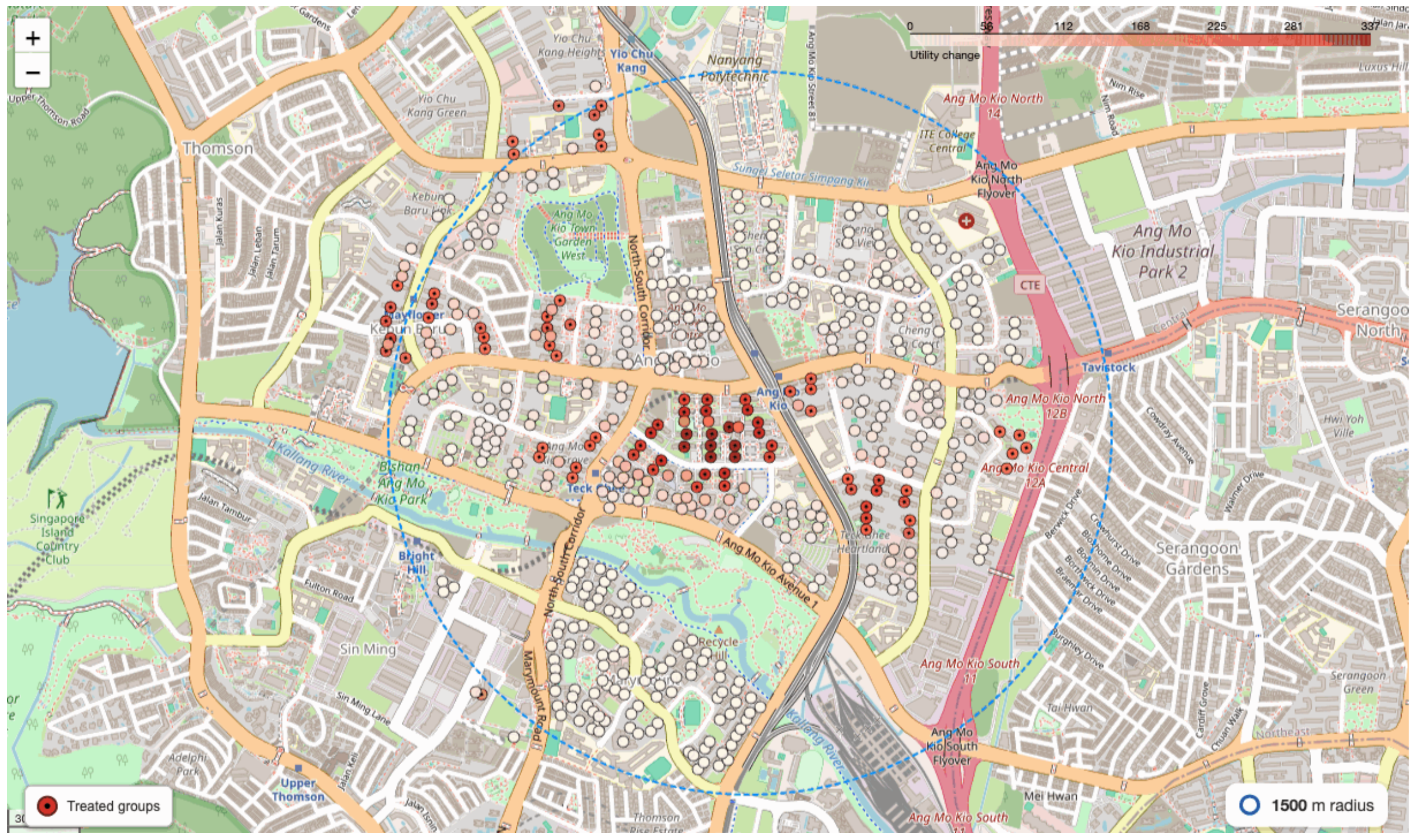

Figure 4: Model-Implied Utility Gains in Ang Mo Kio

**Notes:** This figure maps model-implied utility gains for treated and untreated buildings in Ang Mo Kio. Utility changes are computed relative to the pre-subsidy baseline using the estimated parameters ($\delta_0$ , $\delta_1$ , $\sigma$) from the indirect inference estimation described in Online Appendix Section F. Red (filled) markers denote treated (upgraded) buildings. Open circles denote untreated buildings; the shading intensity reflects the magnitude of utility gain, as indicated by the color scale bar at the top of the figure. The dashed circle denotes a 1,500-meter radius centered on the earliest cohort of treated buildings. Untreated buildings closer to treated sites exhibit larger gains, consistent with distance-decaying housing externalities.

We now turn to the aggregate welfare implications. Table 2 reports welfare changes under the estimated model, comparing scenarios with and without housing externalities. Without externalities, the MUP increases treated households' welfare by 2.43%, with no impact on untreated households (by definition). With externalities, the same subsidy raises treated households' welfare by 3.35%, reflecting both direct benefits from improved housing services and indirect benefits from neighbors' improvements. Untreated neighboring households experience an average welfare gain of 0.11%. Across the entire pop-ulation, the welfare increases by 0.18% and 0.35% in the scenarios without and with externalities. The smaller untreated and full-population averages reflect the large base over which the gains are spread: treated buildings account for only about 7% of baseline welfare, and most untreated buildings lie far from any upgraded site.

Table 2: Welfare Changes from Housing Upgrading and Cash Transfers

| Δ Welfare Relative to Before Subsidy (Without Externality) $(W_p - W)/W$ | | |
|---|---|---|
| | Housing | Cash |
| Treated groups | 2.43% | 2.53% |
| Untreated groups | NA | NA |
| Total population | 0.18% | 0.19% |
| **Δ Welfare Relative to Before Subsidy (With Externality) $(W_p - W)/W$** | | |
| | Housing | Cash |
| Treated groups | 3.35% | 2.36% |
| Untreated groups | 0.11% | 0.01% |
| Total population | 0.35% | 0.18% |

*Notes:* This table reports welfare changes relative to the pre-subsidy baseline, computed as $\frac{W_p - W}{W}$, where $W_p$ denotes welfare under policy $p$ and $W$ denotes baseline welfare. The upper panel reports results for the model equilibrium without housing externalities; the lower panel reports results with externalities. Within each panel, the Housing column reports welfare changes under the in-kind housing subsidy (MUP) and the Cash column under an equivalent lump-sum cash transfer. Rows report results for treated (upgraded) buildings, untreated neighboring buildings, and the full sample. Without externalities, the policy has no spillover effect on untreated buildings by construction (reported as NA). Welfare is computed using the estimated parameters described in Online Appendix Section F.

Housing externalities imply that coordinated upgrading programs can outperform equivalent cash transfers. We compare the MUP to an equivalent lump-sum cash transfer, under which households choose freely how much to allocate to housing upgrading versus other consumption.[13] Without externalities, the cash transfer raises treated households' welfare by 2.53%, exceeding the 2.43% gain under in-kind upgrading due to the flexibility in allocation. With externalities, however, the cash transfer raises treated welfare by only 2.36%, substantially below the 3.35% gain under the MUP. The mechanism is free-riding: under cash transfers, treated households underinvest in housing relative to the social optimum, dampening positive spillovers and reducing aggregate welfare. Untreated neighbors also experience smaller welfare gains under cash transfers.

In Singapore's dense setting, housing externalities are sufficiently strong to offset the distortions from in-kind provision. The welfare ranking of in-kind upgrading over cash transfers, however, depends critically on population density. Table 3 reports counterfactual analyses in which neighborhood density is reduced to approximately 40% of Singapore's level (roughly equivalent to Los Angeles) or to 6% (comparable to Birmingham, Alabama). As density declines, the welfare advantage of housing subsi-

[13]Our cash counterfactual lets the transfer generate externalities only through the housing that households choose to buy with it; it abstracts from any local consumption-amenity externalities, such as more or better neighborhood services, that cash-financed spending on other goods might create. Because this omitted channel would raise the welfare of the cash arm, the in-kind advantage we report in dense settings should be read as an upper bound. We expect the bias to be second-order, however, since consumption-amenity externalities tend to be weaker and more spatially diffuse than the localized housing externalities we estimate, and marginal cash spending is spread across many goods with limited local spillovers.

Table 3: Welfare Changes from Housing Upgrading and Cash Transfers across Density Levels

| $\Delta$ Welfare Relative to Before Subsidy (With Externality) $(W_p - W)/W$ | | | | | | |
|---|---|---|---|---|---|---|
| | Singapore | | 40% (~Los Angeles) | | 6% (~Birmingham) | |
| | Housing | Cash | Housing | Cash | Housing | Cash |
| Treated | 3.35% | 2.36% | 2.78% | 2.43% | 2.48% | 2.51% |
| Untreated | 0.11% | 0.01% | 0.05% | 0.01% | 0.01% | 0.00% |
| Total | 0.35% | 0.18% | 0.25% | 0.19% | 0.19% | 0.19% |

*Notes:* This table reports welfare changes relative to the pre-subsidy baseline across different neighborhood density levels, computed in the counterfactual equilibrium with externalities as $\frac{W_p - W}{W}$, where $W_p$ denotes welfare under policy $p$ and $W$ denotes baseline welfare. For each density scenario, the Housing column reports results for the in-kind housing subsidy and the Cash column for an equivalent lump-sum cash transfer. Rows report results for treated buildings, untreated neighboring buildings, and the full sample. Singapore's population density is approximately 8,207 persons per square kilometer. The counterfactual density levels correspond to approximately 40% of Singapore's density (3,210 per km$^2$, roughly equivalent to Los Angeles) and 6% of Singapore's density (530 per km$^2$, comparable to Birmingham, Alabama).

dies over cash transfers diminishes and eventually reverse: lower density weakens exposure to nearby upgrading and reduces spillover benefits. The welfare ranking of in-kind subsidies versus cash transfers is therefore inherently tied to the density environment. Note that in constructing these counterfactuals we hold the externality parameters fixed and vary only density, which is a conservative choice. The heterogeneity documented in Section 5 indicates that the externality coefficients are themselves smaller in lower-density neighborhoods; allowing the externalities to weaken with density would further reduce the welfare gains from in-kind upgrading in low-density settings, rendering cash transfers an even more clearly welfare-improving alternative.

# 7 Mechanism Analysis

Our baseline estimates document positive, sharply distance-decaying spillovers generated by housing upgrading, but the baseline results alone do not speak to the channels through which upgrading benefits neighbors. The literature points to two broad classes of externalities. The first operates through the physical condition of the local housing stock: its visible quality enters neighbors' valuations directly, so that neglected or distressed buildings depress nearby property values while visible improvements raise them. Because these "eyesore" effects work through physical proximity, they are difficult to measure directly and are typically inferred from their steep decay with distance, as in the literature on foreclosure and mortgage-default externalities (Harding et al., 2009) and on the neighborhood price effects of property rehabilitation (Ganduri and Maturana, 2024). The pronounced distance gradient in our spillover estimates is consistent with this physical channel.

The second class consists of allocative, or compositional, externalities: the characteristics of residents shape local amenities, public goods, and the peer environment, and thereby location desirability and prices. This literature has emphasized sorting along income and skill, whereby upgrading draws in higher-income or higher-skill households whose presence further raises neighborhood amenities and amplifies the initial shock (Guerrieri et al., 2013; Autor et al., 2014). We do not directly observe residents' income or education. As a partial proxy, we construct a wealth measure equal to the value of each resident's housing unit at the beginning of the sample period, and propagate this baseline value forward across subsequent years for the same individual. Because this proxy is noisy and likely attenuates the estimated coefficients, we treat the resulting wealth evidence as suggestive rather than definitive.

We do, by contrast, observe residents' ages directly in administrative records, which gives a cleaner window into compositional change and lets us examine a margin new to this literature. By tracking residential mobility, the evolving age profile, and the wealth proxy of residents, we provide direct, micro-level evidence that upgrading reshapes who lives in treated neighborhoods, offering concrete support for a compositional mechanism underlying the upgrading externalities documented above. Using the administrative residents dataset described in Section 3, we link each resident to the MUP program data by matching buildings and wave years.

Table 4: The Impact of Main Upgrading on Residential Mobility and Resident Age

| | Residential Mobility | | | Resident Age | | |
|---|---|---|---|---|---|---|
| | (1) | (2) | (3) | (4) | (5) | (6) |
| Own Effect – After Announcement and Before Billing | -0.0822*** | -0.1030*** | -0.0968*** | 1.0379*** | 1.1836*** | 1.1901*** |
| | (0.0250) | (0.0230) | (0.0223) | (0.1551) | (0.1449) | (0.1412) |
| Own Effect – After Billing | -0.1224*** | -0.1519*** | -0.1422*** | 1.5907*** | 1.7948*** | 1.8096*** |
| | (0.0362) | (0.0340) | (0.0323) | (0.2217) | (0.2052) | (0.1945) |
| Neighboring Effect (0-500m) – After Announcement | -0.0051*** | -0.0036*** | -0.0035*** | 0.0281*** | 0.0171*** | 0.0153*** |
| | (0.0008) | (0.0007) | (0.0007) | (0.0042) | (0.0041) | (0.0042) |
| Neighboring Effect (500-1000m) – After Announcement | | -0.0034*** | -0.0029*** | | 0.0187*** | 0.0148*** |
| | | (0.0004) | (0.0004) | | (0.0028) | (0.0028) |
| Neighboring Effect (1000-1500m) – After Announcement | | | -0.0022*** | | | 0.0130*** |
| | | | (0.0004) | | | (0.0027) |
| $N$ | 490327 | 646076 | 770836 | 490327 | 646076 | 770836 |
| adj. $R^2$ | 0.224 | 0.217 | 0.184 | 0.067 | 0.065 | 0.062 |
| Controls | Yes | Yes | Yes | Yes | Yes | Yes |
| Expected Exposure | Yes | Yes | Yes | Yes | Yes | Yes |
| Year FE | Yes | Yes | Yes | Yes | Yes | Yes |
| Building FE | Yes | Yes | Yes | Yes | Yes | Yes |

*Notes:* This table reports the effects of the upgrading program on residential mobility (Columns (1)–(3)) and resident age composition (Columns (4)–(6)). For mobility, the dependent variable is an indicator equal to one if an individual's registered residential address differs between the current wave and the subsequent wave, and zero otherwise; for age, the dependent variable is the resident's age at the subsequent wave. The sample consists of individuals observed in the administrative residents dataset across the 1996, 1998, 2000, 2005, and 2011 waves. "Own Effect" captures the direct impact of upgrading on residents of treated buildings, separately for the post-announcement, pre-billing period and the post-billing period. "Neighboring Effect" measures cumulative post-announcement neighborhood exposure (combining both pre-billing and post-billing phases) within the specified distance band. Within each outcome, the first column includes neighboring exposure within 0–500m and the corresponding expected exposure, the second further adds 500–1,000m, and the third additionally includes 1,000–1,500m. All specifications include fixed effects for flat type, storey range, flat model, building, and year. Standard errors clustered at the building level are reported in parentheses. * $p < 0.1$, ** $p < 0.05$, *** $p < 0.01$.

The mechanism analysis is based on the following specification:

$$Mech_{j,t} = \rho_0 + \rho_1 \cdot \mathbb{1}(announce)_{b(i),t} + \rho_2 \cdot \mathbb{1}(billing)_{b(i),t}$$
$$+\rho_3 \cdot Neighbor^{0-500}_{b(i),t} + \rho_4 \cdot Neighbor^{500-1000}_{b(i),t} + \rho_5 \cdot Neighbor^{1000-1500}_{b(i),t}$$
$$+ + \rho_6 \cdot Neighbor^{0-500}_{b(i),t} + \rho_7 \cdot Neighbor^{500-1000}_{b(i),t} + \rho_8 \cdot Neighbor^{1000-1500}_{b(i),t} +$$
$$+X_{i,t}{}' \cdot \rho_9 + \alpha_{b(i)} + \alpha_t + \xi_{j,t} \tag{3}$$

where i denotes the housing unit occupied by resident j at wave t, b(i ) its building, and t the wave year. We consider two outcomes, $Mech_{j,t}$ a mobility indicator equal to one if the individual's registered address differs between wave t and the subsequent wave, and zero otherwise; and (ii) the resident's age at the subsequent wave. The treatment-phase indicators follow the definitions in Section 3.2. The variable $Neighbor^{r}_{b(i),t}$ measures cumulative post-announcement neighborhood exposure (combining both pre-billing and post-billing phases) within distance band r, defined as in Equation (2), and $\overline{\text{Neighbor}}^{r}_{b(i),t}$

is the corresponding expected exposure. We control for flat characteristics Xi,t (flat type, storey range, and flat model fixed effects), building fixed effects $\alpha$b(i), and year fixed effects $\alpha_t$.

Columns (1)–(3) of Table 4 report the effects of upgrading on mobility. The own effects are negative and precisely estimated in both the post-announcement and post-billing periods, indicating that residents in treated buildings become substantially less likely to move. Spillovers are also present: residents near upgraded buildings are less likely to relocate, with neighboring effects that are negative, statistically significant, and attenuating with distance (0–500m, 500–1,000m, and 1,000–1,500m). This spatial decay pattern is consistent with upgrading improving local housing services and neighborhood conditions, increasing place attachment not only for treated households but also for nearby non-recipients. These results support a mobility-based mechanism: upgrading raises local amenity value, increases household retention, and lowers residential turnover in treated areas and surrounding neighborhoods. Consistent with reduced turnover, resale transaction volume also declines: Online Appendix Table C.13 reports negative, distance-decaying neighboring effects on the number of resale transactions, mirroring the mobility results.

If reduced mobility occurs disproportionately within certain subgroups, and if heterogeneous residents generate group-specific neighborhood amenities, such differential retention will also produce the spatial decay pattern observed in price capitalization. In the context of upgrading aged housing buildings, resident age composition serves as a natural proxy: if upgrading disproportionately increases place attachment for older households and motivates them to stay, the neighborhood age profile will shift upward. Consistent with this prediction, Columns (4)–(6) of Table 4 show that the average age of residents rises in treated buildings in both the post-announcement and post-billing periods, and that nearby buildings also experience positive, statistically significant increases with a spatial gradient, supporting differential retention as the operative mechanism. Because neighbor demographic characteristics influence both local amenity provision and social interactions, these compositional shifts toward middle- and older-aged residents (with an average age of 46) provide evidence of an allocative externality channel that can operate alongside, and potentially reinforce, physical investment improvements (Autor et al., 2014; Guerrieri et al., 2013).

Turning to the wealth proxy, Online Appendix Table C.14 shows that upgrading is associated with a decline in the average wealth of residents in treated buildings, while the average wealth of residents in nearby buildings is essentially unchanged. We are cautious about pushing these estimates too far given the noisiness of the proxy, which likely attenuates the coefficients. Read alongside the mobility and age findings, however, the own-building decline is qualitatively consistent with differential retention: upgrading disproportionately retains older incumbents, who may tend to be relatively wealth-poorer, so the lower average wealth at the treated-building level reflects compositional change in who stays, but the externality channel is not clear from this proxy.

# 8 Conclusion

Housing availability is one of the defining challenges of urbanization. Across cities in both the developed and developing world, rapid population growth, constrained land supply, and rising construction costs have intensified demand for policy interventions that improve available housing services. In-kind housing upgrading programs are a prominent response, yet our understanding of how such programs operate in the dense urban environments where they are most needed has been limited. This paper provides new evidence from Singapore, where the upgrading programme was implemented across one of the world's densest residential landscapes.

We show that large-scale housing upgrading in a high-density city generates substantial spillovers that extend beyond directly treated households. Exploiting the staggered implementation of Singapore's MUP and administrative micro data, we find that upgrading increases the value of treated units and raises prices in surrounding neighborhoods. These externalities are highly localized, decaying rapidly with distance. Upgrading also reduces residential mobility in treated buildings, with spatial spillovers that similarly attenuate with distance, and treated neighborhoods exhibit a shift toward an older resident age profile. Together, these results indicate that physical improvements strengthen place attachment and lower residential turnover, reshaping neighborhood composition; these compositional shifts likely operate alongside the physical channel in generating the housing externalities.

To inform policy design, we estimate a simple spatial housing-services model using moments from our reduced-form estimates and quantify the welfare trade-off between in-kind upgrading subsidies and equivalent lump-sum cash transfers. In dense cities such as Singapore, spillovers are sufficiently strong that upgrading yields larger welfare gains than cash transfers once externalities are taken into account. In lower-density counterfactuals, however, spillovers weaken and the welfare advantage of in-kind upgrading diminishes, with cash transfers potentially dominating in sufficiently sparse settings. These findings underscore that the optimal form of housing subsidy is inherently density-dependent and that welfare evaluations of upgrading policies should explicitly account for the extent of housing externalities.

Several directions for future research emerge from this analysis. First, our welfare model treats households as homogeneous in preferences; incorporating heterogeneity would allow the analysis to capture distributional consequences of upgrading and the potential for gentrification-induced displacement. Second, while we document compositional shifts through residential mobility and age, richer data on household income and education would permit a more direct decomposition of different types of allocative externalities. Third, our model abstracts from general equilibrium price adjustments in the broader housing market; embedding the externality structure in a spatial equilibrium framework with endogenous location choice would enable a more complete welfare accounting. Finally, extending the analysis to other dense urban contexts with large-scale housing programs would help establish the external validity of the density-dependent welfare ranking.